\documentclass[aps,prc,reprint,superscriptaddress, showpacs]{revtex4-1}

\usepackage{mathptmx}

\usepackage{pstricks}
\usepackage{array}
\usepackage{tabularx}
\usepackage{graphicx}
\usepackage{subfigure}
\usepackage{color}
\usepackage{filemod}
\usepackage{psfrag}
\usepackage{amsmath}
\usepackage{amssymb}
\usepackage{psfrag}
\usepackage{textcomp}
\usepackage{longtable}
\usepackage{multirow}
\usepackage{braket}
\usepackage[mathscr]{euscript}
\usepackage{dcolumn}
\usepackage{bm}
\usepackage{hyperref}
\hypersetup{letterpaper, colorlinks=true, urlcolor=blue, citecolor=blue, linkcolor=blue, breaklinks=true}
\usepackage{breakurl}

\newcommand{\msu}{Department of Physics and Astronomy, Michigan State University, East Lansing, Michigan 48824, USA}
\newcommand{\msuchem}{Department of Chemistry, Michigan State University, East Lansing, Michigan 48824, USA
}
\newcommand{\nscl}{National Superconducting Cyclotron Laboratory, Michigan State University, East Lansing, Michigan 48824, USA}
\newcommand{\augustana}{Department of Physics and Astronomy, Augustana College, Rock Island, Illinois 61201, USA}

\newcommand{\hope}{Department of Physics, Hope College, Holland, Michigan 49423, USA}
\newcommand{\cmu}{Department of Physics, Central Michigan University, Mt. Pleasant, Michigan 48859, USA}

\newcommand{\triumf}{TRIUMF, 4004 Wesbrook Mall, Vancouver, British Columbia V6T 2A3, Canada}
\newcommand{\cord}{Department of Physics, Concordia College, Moorhead, Minnesota 56562, USA}

\newcommand{\ganil}{GANIL, CEA/DSM-CNRS/IN2P3, Bvd Henri Becquerel, 14076 Caen, France}

\newcommand{\nuc}[2]{$^{#1}\textrm{#2}$}
\newcommand{\figref}[1]{Fig.~\ref{#1}}
\newcommand{\eqnref}[1]{Eq.~\ref{#1}}
\newcommand{\tableref}[1]{Table~\ref{#1}}

\newcommand{\secref}[1]{Section~\ref{#1}}

\newcommand{\executeiffilenewer}[3]{%
\if{\Filemodnewest[1]{#1}{#2} == #2}
{\immediate\write18{#3}}\fi%
}

\graphicspath{{./figures/}}

\begin{document}

\title{Spectroscopy of Neutron-Unbound $^{27,28}\textrm{F}$}

\author{G. Christian}
\thanks{Present address: \triumf. \\ gchristian@triumf.ca}
\affiliation{\msu}
\affiliation{\nscl}

\author{N. Frank}
\affiliation {\augustana}

\author{S. Ash}
\affiliation {\augustana}

\author{T. Baumann}
\affiliation {\nscl}

\author{P. A. DeYoung}
\affiliation{\hope}

\author{J. E. Finck}
\affiliation{\cmu}

\author{A. Gade}
\affiliation{\msu}
\affiliation{\nscl}

\author{G. F. Grinyer}
\thanks{Present address: \ganil.}
\affiliation{\nscl}

\author{B. Luther}
\affiliation{\cord}

\author{M. Mosby}
\affiliation{\nscl}
\affiliation{\msuchem}

\author{S. Mosby}
\affiliation{\msu}
\affiliation{\nscl}

\author{J. K. Smith}
\affiliation{\msu}
\affiliation{\nscl}

\author{J. Snyder}
\affiliation{\msu}
\affiliation{\nscl}

\author{A. Spyrou}
\affiliation {\msu}
\affiliation {\nscl}

\author{M. J. Strongman}
\affiliation{\msu}
\affiliation{\nscl}

\author{M. Thoennessen}
\affiliation{\msu}
\affiliation{\nscl}

\author{M. Warren}
\affiliation {\augustana}

\author{D. Weisshaar}
\affiliation{\nscl}

\author{A. Wersal}
\affiliation{\nscl}

\begin{abstract}

The ground state of \nuc{28}{F} has been observed as an unbound resonance $2\underline{2}0$ keV above the ground state of $^{27}$F. Comparison of this result with USDA/USDB shell model predictions leads to the conclusion that the $^{28}$F ground state is primarily dominated by $sd$-shell configurations. Here we present a detailed report on the experiment in which the ground state resonance of $^{28}$F was first observed. Additionally, we report the first observation of a neutron-unbound excited state in $^{27}$F at an excitation energy of $25\underline{0}0 (2\underline{2}0)$ keV.

\end{abstract}


\maketitle

\section{Introduction}

The neutron-rich region around $N = 20$ has been a topic of active experimental and theoretical research for over $30$ years, owing to the transition from pure $sd$ to mixed $sd$-$pf$ shell configurations first deduced from mass measurements of neutron-rich sodium isotopes \cite{PhysRevC.12.644, PhysRevC.41.1147}.  As the available intensities at rare-isotope beam facilities have increased, it has become possible to explore increasingly neutron-rich systems near $N = 20,$ including those at and beyond the neutron dripline.

The heavy fluorine isotopes represent some of the most neutron-rich $N \sim 20$ systems that can be measured with present experimental techniques.  A 2004 measurement \cite{Elekes200417} reported two $\gamma$-ray transitions in each of the $^{25,26,27}\textrm{F}$ isotopes, with the higher-lying \nuc{26}{F} transition confirmed in a recent experiment \cite{PhysRevC.85.017303}.  In \nuc{27}{F}, a transition was observed at  $777(19)$ keV and assigned to the first $1/2^{+}$ excited state. This is in poor agreement with USD \cite{USD1, USD2} shell model predictions which place the $1/2^{+}$ at $1997$ keV. SDPF-M Monte Carlo Shell Model calculations \cite{PhysRevC.60.054315}, which allow for $sd$-$pf$ shell mixing, are in better agreement with observation, placing the $1/2^{+}$ at $1100$ keV.   This suggests that the first $1/2^{+}$ excited state in \nuc{27}{F} exhibits significant $sd$-$pf$ configuration mixing. Additionally, Ref. \cite{Elekes200417} reports a low-energy transition not predicted by USD in each of the \nuc{25,26,27}{F} isotopes, speculating that these transitions might correspond to $1/2^{-}$ states arising from proton $p$-$sd$ cross shell excitations.

Until recently only one measurement of neutron-unbound states in fluorine isotopes has been reported. A $28$ keV resonant decay from \nuc{25}{F} was assigned to a $1/2^{-}$ excited state in \nuc{25}{F} at an excitation energy of $4249$ keV \cite{PhysRevC.84.037302}. In the present paper we report the observation of the unbound ground state of \nuc{28}{F} and an unbound excited state in \nuc{27}{F}. The results of the \nuc{28}{F} experiment have been reported in a recent article \cite{PhysRevLett.108.032501}.
\section{Experimental Details}
\subsection{Setup}

\begin{figure*}
    \centering
    \includegraphics{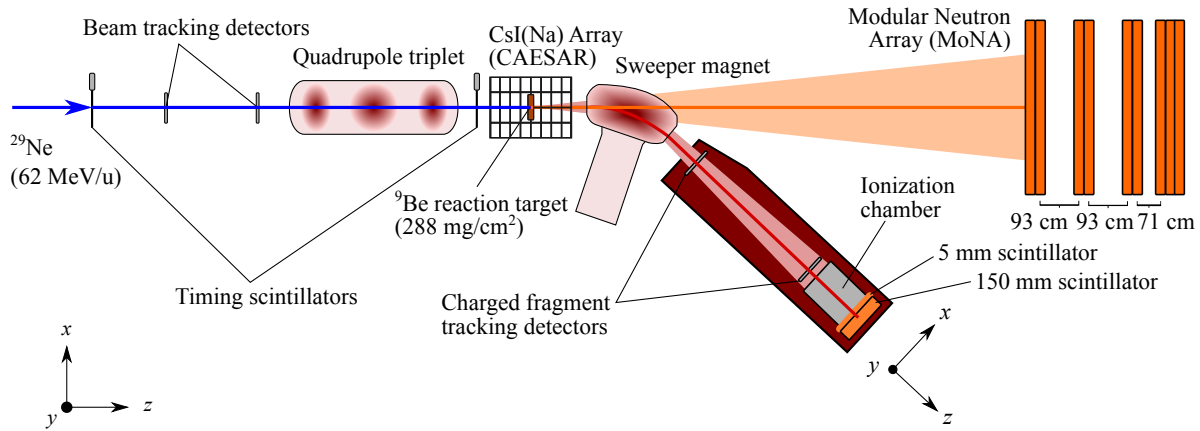}
	\caption{(color online) Diagram of the experimental setup.}
	\label{fig:setup}
\end{figure*}

\begin{figure}
    \centering
	\includegraphics{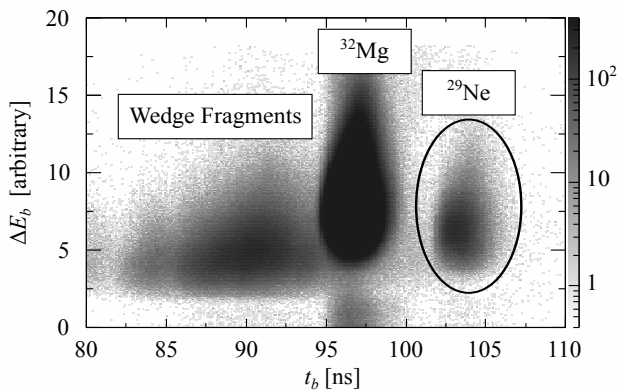}
	\caption{Energy loss versus flight time of the secondary beam, showing its three components: \nuc{29}{Ne}, \nuc{32}{Mg}, and various lighter species (``wedge fragments'') produced in the A1900 wedge. }
	\label{fig:beamid}
\end{figure}

The experiment was performed at the National Superconducting Cyclotron Laboratory (NSCL) at Michigan State University. Unbound states in \nuc{27,28}{F} were populated by nucleon removal from a beam of \nuc{29}{Ne}.  The \nuc{29}{Ne} beam was produced by first accelerating $^{48}\textrm{Ca}^{20+}$ to $140$ MeV/u in the NSCL coupled cyclotrons \cite{marti:64}. The \nuc{48}{Ca} then impinged upon a $1316$ $\textrm{mg/cm}^2$ \nuc{9}{Be} production target.  Products of the calcium on beryllium reaction were sent through the A1900 fragment separator \cite{Morrissey200390}, which was tuned to optimize the transmission of \nuc{29}{Ne} at $62$ MeV/u.  The A1900 included an achromatic aluminum wedge at its second image point to disperse fragments according to $A/Z$ and improve separation.

After the A1900, a quadrupole triplet magnet focused the beam onto a $288$ mg/cm$^2$ \nuc{9}{Be} reaction target.  Upstream of the target, the beam passed through a pair of position-sensitive cathode readout drift chambers (CRDCs) separated by $227~${}cm and a pair of plastic scintillators separated by $1044~${}cm. The location of each of these detectors along the beam axis is shown in \figref{fig:setup}. The CRDC position measurements were used to calculate the beam position on the reaction target by ray tracing through the quadrupole triplet. The upstream scintillator was $1010~${}$\mu${}m thick, and the downstream (``target'') scintillator was $254~${}$\mu${}m. Each scintillator recorded a time signal, and these signals were used to calculate the beam time of flight $(t_{b})$.  Additionally, the target scintillator recorded an energy loss signal $(\Delta E_{b})$.  As shown in \figref{fig:beamid}, the various beam components were well separated in energy loss versus time of flight. The desired \nuc{29}{Ne} composed approximately $2 \%$ of the beam, and the remainder was composed of \nuc{32}{Mg} $(87 \%)$ and various lighter species produced in the aluminum wedge.  

A diagram of the experimental setup is shown in \figref{fig:setup}. The experiment consisted of three subsystems, each used to measure a different type of reaction residue potentially resulting from the breakup of neutron-unbound states in \nuc{27,28}{F}: neutrons, $\gamma$ rays (from feeding to bound excited states in the daughter), and residual charged particles.  Neutrons were detected in the Modular Neutron Array (MoNA) \cite{Baumann2005517}, which measured their time of flight, position, and the amount of light deposited.  $\gamma$ Rays were detected in the Caesium Iodide Array (CAESAR) \cite{Weisshaar2010615}, which measured their total energy and time of flight.  Charged particles were first deflected $43^\circ$ by the Sweeper magnet \cite{1439869}. They were then detected in a pair of CRDCs, an ionization chamber measuring energy loss, and two plastic scintillators. The front face of each scintillator was $40~\textrm{cm} \times 40~\textrm{cm},$ and each was coupled to four photo-tubes.  The upstream scintillator was $5$ mm thick and recorded a time signal; this signal was combined with the time output of the target scintillator to determine the fragment time of flight.  The downstream scintillator was $150$ mm thick, and its charge output was indicative of the fragment energy.

Due to the size and complexity of the setup, separate data acquisition (DAQ) systems were used for MoNA and Sweeper-CAESAR.  Events in each DAQ were recorded with a timestamp, allowing coincidences to be reconstructed off-line. Although run separately, the triggering of each DAQ was controlled by a shared logic module, which allowed for trigger conditions involving both subsystems. To reduce dead-time, the experiment required coincidences between MoNA and the $5$ mm scintillator located at the back of the Sweeper box. CAESAR detected $\gamma$ rays in coincidence, but they were not a required trigger condition.

\subsection{Data Analysis}

\subsubsection{Charged Particle Separation}

\begin{figure*}
    \centering
   	\subfigure	{\includegraphics{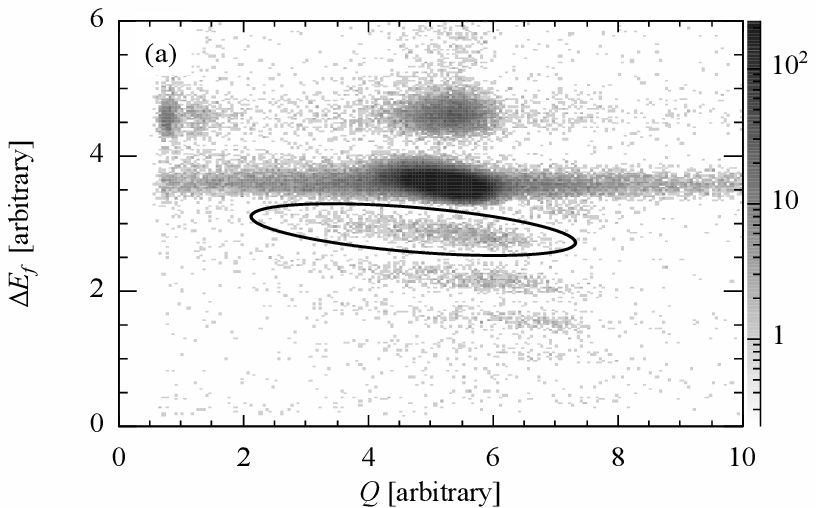}
   			\label{fig:eid_thick}
   				}
    \subfigure	{\includegraphics{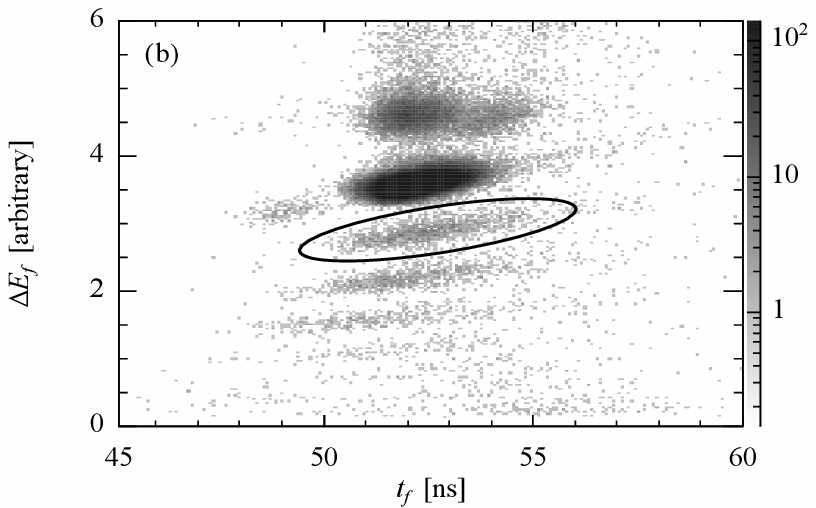}
      		\label{fig:eid_TOF}
      			}
    \subfigure	{\includegraphics{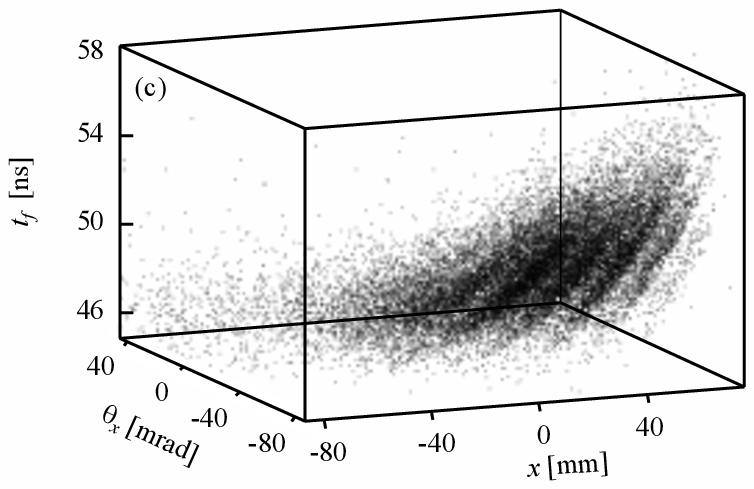}
      		\label{fig:pid3d}
      			}
    \subfigure	{\includegraphics{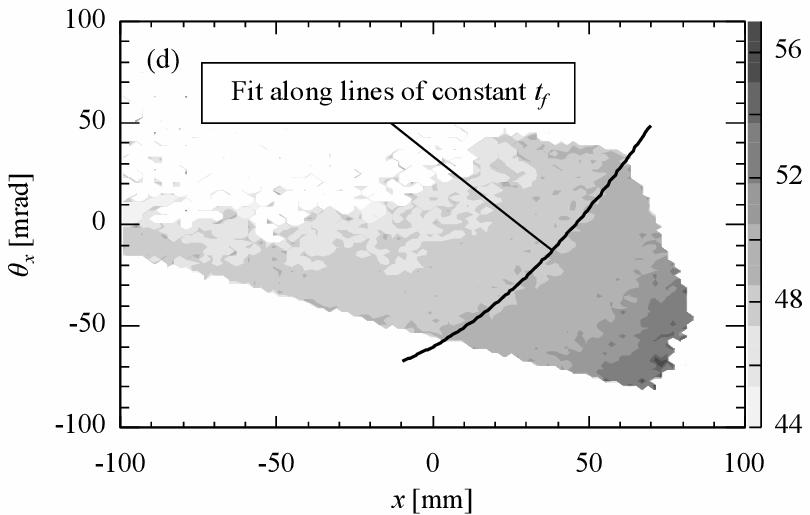}
      		\label{fig:pid_profile}
      			}
    \subfigure	{\includegraphics{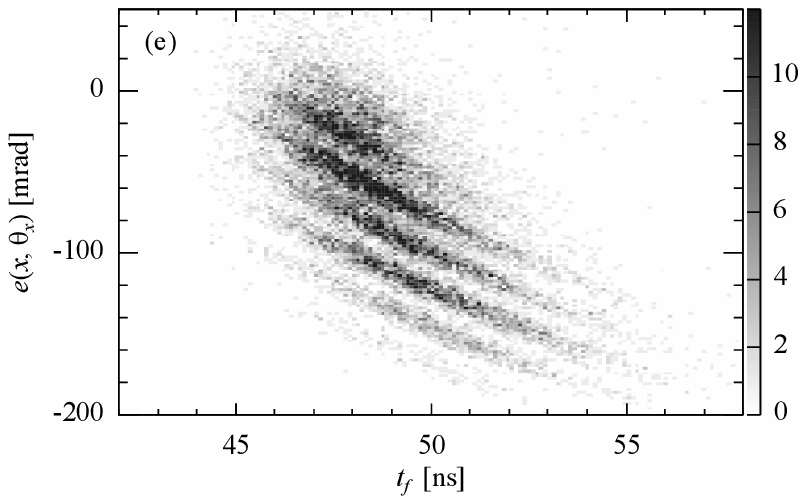}
      		\label{fig:pid_xtx}
      			}
    \subfigure	{\includegraphics{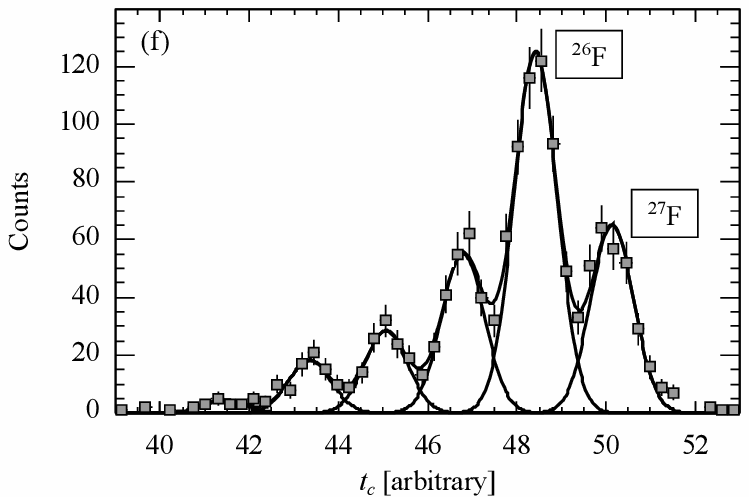}
      		\label{fig:pid_final}
      			}

	\caption{a) Element separation in $\Delta E$-$E.$ b) Element separation in $\Delta E$-$t_{b}$.  Panels (a) and (b) are both gated on incoming \nuc{29}{Ne} and have the fluorine events circled.  c) Three-dimensional plot of $t_{b}$ vs. dispersive position and angle after the Sweeper for fluorine fragments produced from the \nuc{32}{Mg} beam. The various bands correspond to different isotopes of fluorine. d) Profile of the three-dimensional plot in (c), including the emittance parameter, $e(x, \theta_x),$ that is determined by fitting lines of constant $t_{b}$ with a second order polynomial. e)  Plot of the $e(x, \theta_x)$ parameter determined in (c) versus $t_{b}$, demonstrating isotope bands in two dimensions.  f) Final corrected time of flight $(t_{c})$ for fluorine elements produced from the \nuc{29}{Ne} beam, with \nuc{26}{F} and \nuc{27}{F} indicated.}
	\label{fig:pid}
\end{figure*}

The charged particle measurements allowed for event-by-event isotope identification after making a variety of corrections to the data.  The first step was to identify the various elements reaching the end of the Sweeper using measurements of energy loss and total energy.  Energy loss $(\Delta E_f)$ was obtained from the ionization chamber signal. The fragment time of flight $(t_{f})$ and charge output of the $150~${}mm scintillator $(Q)$ were each used as an independent indicator of total energy.  Figures \ref{fig:eid_thick} and \ref{fig:eid_TOF} show the element separation in $\Delta E_f$-$Q$ and $\Delta E_f$-$t_{f}$, respectively. In the final analysis, events were required to fulfill conditions in both parameter spaces.

For a given element, isotopes were separated by constructing a corrected time of flight parameter $(t_{c})$ indicative of $A/Z$.  The corrections to the time of flight accounted for the varying paths taken through the Sweeper, and the primary indicators of this path length were the dispersive position ($x$) and dispersive angle $(\theta_x)$ of the fragment as it exited the magnet. Additionally, a variety of other parameters (c.f. \tableref{table:tcorr}) were found to correlate with the time of flight of a given isotope and were included in the corrections.  Due to the lack of focusing elements, as well as non-homogeneities in the Sweeper's magnetic field, it was necessary to consider three-dimensional correlations between $t_{f}$, $x$, and $\theta_x$, as well as non-linearities, in determining the appropriate corrections.  Because of superior statistics, the time of flight corrections were determined for fluorine elements produced from the \nuc{32}{Mg} beam. These same corrections were then used to separate the isotopes of interest, fluorines produced from \nuc{29}{Ne}.

\begin{table}
\caption{Correction factors used for isotope separation. To calculate the corrected time of flight, we take the sum of each factor multiplied by its corresponding parameter and then add this sum to $t_\textit{f}$. The symbols $x (y)$ and $\theta_x (\theta_y)$ respectively refer to the dispersive (non-dispersive) position and angle of the charged fragment as it exits the Sweeper. The symbol $x_\mathit{trgt} (y_\mathit{trgt})$ denotes the beam's dispersive (non-dispersive) position on the reaction target. The remaining symbols are all introduced in the text.}
	\centering
	\begin{ruledtabular}
	\begin{tabular}{ll}
		\textit{Parameter}												& \textit{Correction Factor}   \\
		\hline \\[-8pt]
		$x                             	$ (mm)						&  $-5.0595 \times 10^{-2}$    \\
		$x^2                           	$ (mm$^{2}$)				&  $-8.97\times 10^{-4}$       \\
		$x^3                           	$ (mm$^{3}$)				&  $-3.0 \times 10^{-6}$       \\
		$\theta_x                      	$ (mrad)					&  $+8.0 \times 10^{-2}$       \\
		$\theta_x^2                    	$ (mrad$^{2}$)				&  $-1.0 \times 10^{-5}$       \\
		$\theta_x^3                    	$ (mrad$^{3}$)				&  $+2.0 \times 10^{-6}$       \\
		$x \theta_x                    	$ (mm mrad)					&  $-1.5 \times 10^{-4}$       \\
		$x \theta_x^2                  	$ (mm mrad$^{2})$			&  $-6.0 \times 10^{-6}$       \\
		$x^2 \theta_x                  	$ (mm$^{2}$ mrad)			&  $-2.0 \times 10^{-6}$       \\
		$x^2 \theta_x^2                	$ (mm$^{2}$ mrad$^{2}$)	&  $+1.4 \times 10^{-7}$       \\
		$y^2                           	$ (mm$^{2}$)				&  $+1.0 \times 10^{-3}$       \\
		$\theta_y                      	$ (mrad)					&  $-3.0 \times 10^{-3}$       \\
		$x_\mathit{trgt}	       			$ (mm)						&  $+1.7 \times 10^{-2}$       \\
		$y_\mathit{trgt}	              	$ (mm)						&  $+4.0 \times 10^{-3}$       \\
		$t_\mathit{b}	  				  	$ (ns)						&  $+1.0 \times 10^{-1}$       \\
		$Q					             	$ (arb.)					&  $+1.3 \times 10^{-3}$       \\
		$\Delta E							$ (arb.)					&  $+4.0 \times 10^{-3}$       \\
	\end{tabular}
	\end{ruledtabular}
\label{table:tcorr}
\end{table}

As shown in \figref{fig:pid3d}, a three-dimensional plot of $t_{f}$-$x$-$\theta_x$ displays isotope bands. For the purpose of time of flight corrections, it is useful to reduce the $x$-$\theta_x$ phase space into a single ``emittance'' parameter $e(x, \theta_x),$ as this will allow for corrections to the flight time to be made in a straightforward way.  To determine $e(x, \theta_x),$ the $t_{f}$-$x$-$\theta_x$ scatter-graph was profiled by dividing the $x$-$\theta_x$ phase space into small regular rectangular regions and finding the mean $t_{f}$ for each region \footnote{In practice, this was done using the \texttt{TH3::Project3DProfile} method of the ROOT data analysis package. For more information, see Chapter 3 of the ROOT users guide: \url{http://root.cern.ch/download/doc/3Histograms.pdf}}.  This profile plot is shown in \figref{fig:pid_profile}, with the grayscale level representing mean $t_{f}$.  From here, the location of $\theta_x$ as a function of $x$ was fit along the lines of constant $t_{f}$ in the profile.  As shown by the curve in \figref{fig:pid_profile}, the location of these lines was well-described by a second order polynomial,

\begin{equation}
\label{eq:isoTOF}
     f \left(x\right) = a x^2 + b x + c~[\textrm{mrad}], 
\end{equation}

\noindent  with $a = $ $0.010391~\textrm{mrad}/\textrm{mm}^2$ and $b = $ $0.84215~\textrm{mrad}/\textrm{mm}$. The final constant $c$ can take on any value; it only causes the curve to shift to a different line of constant $t_{f}$.

Once $f(x)$ was determined, $e(x, \theta_x)$ was constructed simply as

\begin{equation}
\label{eq:emit}
	e(x, \theta_x) = \theta_x - f(x).
\end{equation}

\noindent As shown in \figref{fig:pid_xtx}, plotting $e(x, \theta_x)$ versus $t_{f}$ reveals isotope bands in two dimensions.  From here, an initial corrected time of flight parameter was calculated by projecting onto the axis perpendicular to the bands.  The time of flight corrections were then further refined by iteratively removing any correlations between $t_{f}$ and the parameters listed in \tableref{table:tcorr}.

The final corrected time of flight $(t_{c})$ for fluorines produced from \nuc{29}{Ne} is shown in \figref{fig:pid_final}. By fitting this spectrum with the sum of five Gaussians constrained to have equal width, we determined the \nuc{26}{F}-\nuc{27}{F} cross-contamination to be approximately $4 \%.$ The factors used in constructing the corrected time of flight are listed in \tableref{table:tcorr}, and it should be noted that the most important corrections (in addition to those for $x$, $\theta_x$, and their higher order combinations) are those for $y^2$ and $x_\textit{trgt}.$

\subsubsection{Decay Energy Calculation}
The decay energy of the breakup of unbound states was calculated using invariant mass analysis.  In Euclidian coordinates, the decay energy $E_d$ is expressed as
\begin{equation}
\label{eq:edecay}
	E_{d} =
		\sqrt{m_f^2 + m_n^2 + 2 \left(E_f E_n  - p_f p_n \cos{\theta} \right)}  - m_f - m_n,
\end{equation}

\noindent where $m_{f} (m_{n}),$ $E_{f} (E_{n}),$ and $p_{f} (p_{n})$ refer to the mass, energy, and momentum of the charged fragment (neutron), respectively, and $\theta$ is the opening angle between the two decay products. Charged fragment inputs to \eqnref{eq:edecay} were determined using a partially inverted COSY transformation matrix \cite{Makino2006346, Frank20071478}, which operated on the measured position and angle behind the Sweeper and the $x$ position of the beam on target. The transformation returned the energy and angle at the reaction target, as well as the track length and the target $y$ position.

The neutron input to \eqnref{eq:edecay} was calculated from time of flight and position measurements in MoNA using relativistic kinematics. The trigger logic was designed such that the stop for each MoNA time digitizer channel was provided by a delayed signal from the target scintillator.  Thus the recorded time signals were a measurement of neutron time of flight $(t_n)$. To calibrate the raw digitizer signals, a linear slope and offset were applied to each channel.  The slopes were determined from a pulser run; relative offsets between MoNA bars were determined from cosmic-ray muon tracks; and an overall offset was set from the travel time of prompt $\gamma$ rays. Vertical and lateral positions in MoNA were assumed to be at the center of the interaction bar, and the horizontal position was calculated from the time difference between signals measured on either end of the bar.  In the case of multiple interactions within MoNA, the earliest hit with $t_{n} > 40$ ns was used in the analysis.  The cutoff of $40$ ns was chosen to eliminate any random first hits that arrived too early to be prompt neutrons.

\begin{figure}
    \centering
	\includegraphics{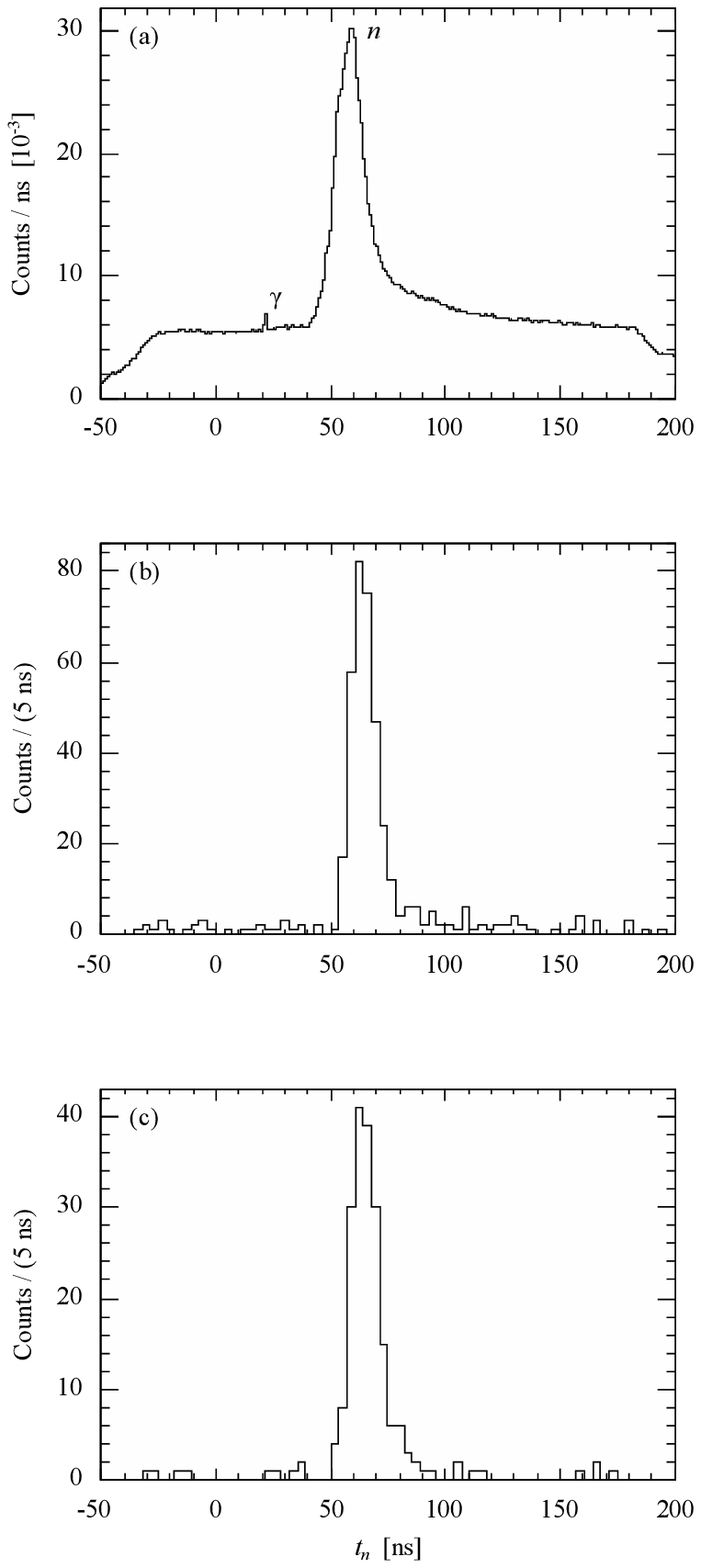}
	\caption{Neutron time of flight to the front face of MoNA, for the following conditions: a) ungated; b) in coincidence with \nuc{26}{F}; c) in coincidence with \nuc{27}{F}.}
	\label{fig:nTOF}
\end{figure}

A plot of the neutron time of flight to the front face of MoNA is presented in \figref{fig:nTOF}, for three conditions: ungated (including all incoming beam components), \nuc{26}{F} produced from \nuc{29}{Ne}, and \nuc{27}{F} produced from \nuc{29}{Ne}.  In the ungated plot, the peaks from prompt neutrons and $\gamma$ rays are clearly identifiable on top of a random flat background consisting primarily of room background $\gamma$ rays and cosmic-ray muons.  When requiring coincidences with \nuc{26,27}{F}, the flat background is essentially eliminated, and the prompt neutron peak dominates the spectrum.

\subsubsection{$\gamma$-Ray Measurements}
CAESAR was calibrated using a variety of standard $\gamma$-ray sources \footnote{The sources and their respective $\gamma$-ray lines (in keV) were: \nuc{133}{Ba} (356), \nuc{137}{Cs} (662), \nuc{22}{Na} (1275), \nuc{88}{Y} (898, 1836), \nuc{56}{Co} (517, 846, 1771, 2034, 2598, 3272).}.  Although a large magnetic shield was placed between it and the Sweeper, CAESAR was still subject to significant fringe fields (on the order of $3$ mT) which affected the response of its photo-tubes.  To account for this, the array was calibrated with the Sweeper set to the rigidity at which the experiment was performed.  Furthermore, to account for potential hysteresis effects, a recalibration run using a \nuc{88}{Y} source was taken any time the field of the Sweeper was changed during the experiment.

For $\gamma$-ray events depositing light in multiple crystals, the total deposited energy was calculated using an ``add-back'' technique \cite{Weisshaar2010615}. The in-beam data were Doppler corrected; for the correction, the detection point taken as the center of the first interaction crystal, and the emission point assumed to be the center of the reaction target.  To reduce background from random coincidences, only events falling within a specific time window were included in the final analysis. Because of electronic effects (walk in the leading-edge discriminators used for timing), the time window was implemented as a two-dimensional cut on time of flight versus Doppler-corrected energy.

\subsection{Modeling and Simulation}
\label{sec:modelsim}

Resonant states were modeled by a Breit-Wigner line-shape with an energy dependent width derived from $R$-Matrix theory \cite{RevModPhys.30.257}. The equation for the line-shape is
\begin{equation}
\label{eq:breit_wigner}
	\sigma(E; E_0, \Gamma_0, \ell)
		=        \frac{A \Gamma_{\ell} \left(E; \Gamma_0 \right)}
			     {\left[ E_0 + \Delta_{\ell} \left( E; \Gamma_0 \right) - E \right]^{2} +
			     \frac{1}{4} \left[\Gamma_{\ell}\left(E; \Gamma_0 \right) \right]^2},
\end{equation}

\noindent where $A$ is an amplitude, $E_0$ is the central resonance energy, $\Gamma_0$ parameterizes the central resonance width, $\ell$ is the orbital angular momentum of the resonance, and $\Gamma_\ell$ and $\Delta_\ell$ are given by

\begin{equation}
\label{eq:gamma_delta}
	\begin{array}{l}
		\Gamma_\ell \left(E\right) = 2 P_\ell \left(E\right)\gamma_0^2 \\[6pt]
		\Delta_\ell \left(E\right) = -\left[S_\ell \left(E\right) -
										   S_\ell \left(E_0\right)
								    \right] \gamma_0^2.
	\end{array}
\end{equation}

The $P_\ell$ and $S_\ell$ functions in \eqnref{eq:gamma_delta} are related to the spherical Bessel Functions, $J_\ell (\rho),$ and their derivatives:
\begin{equation}
\label{eq:shift_pen}
	\begin{array}{l}
		P = \left[ \rho / \left( F^2_\ell + G^2_\ell \right) \right]_{r = a} \\[6pt]
		S = \left[\rho \left(F_\ell F^\prime_\ell + G_\ell G^\prime_\ell \right) /
		            \left(F^2_\ell + G^2_\ell \right) \right]_{r = a} ,

	\end{array}
\end{equation}

\noindent with $F_\ell = \left( \pi \rho / 2 \right)^{1/2} \linebreak[1] J_{\ell+1/2} \left(\rho\right)$ and $G_\ell = \left(-1\right)^\ell \left(\pi \rho /2\right)^{1/2} \linebreak[1] J_{-\left(\ell+1/2\right)} \left(\rho\right).$

In addition to resonant states, a non-resonant background is expected in the $^{27}\textrm{F} \rightarrow ^{26}\textrm{F} + n$ decay energy spectrum, resulting from the decay (via emission of a neutron with $E_{d} \lesssim 3$ MeV) of high-lying continuum states in \nuc{28}{F} to high-lying states in \nuc{27}{F} that subsequently feed the ground state of \nuc{26}{F}. The \nuc{26}{F} fragment can then be detected in coincidence with the first neutron, giving rise to the background distribution.  This background was modeled as a Maxwellian distribution of beam velocity neutrons,
\begin{equation}
\label{eq:therm_final}
	f\left(\epsilon; \Theta \right) =
		A \sqrt{\epsilon / \Theta^3} e^{-\epsilon / \Theta},
\end{equation}

\noindent with the temperature $\Theta$ a free parameter.  This model provides a good fit to the observed non-resonant data and has been employed in number of other invariant mass measurements, for example \cite{Deak198767, PhysRevLett.99.112501, PhysRevC.83.031303, PhysRevLett.100.152502, Spyrou2010129}.

Broadening due to experimental resolution and acceptance was accounted for in a Monte Carlo simulation of the experiment.  In the simulation, the kinetic energy of the incoming \nuc{29}{Ne} beam was modeled as a Gaussian with $E_0 = 62.1~${}MeV/u and $\sigma_E = 1.72~${}MeV/u, clipped at $E < 64.5~${}MeV/u.  The beam angle and position were also modeled as Gaussian with $\sigma_x = 11~${}mm, $\sigma_{\theta x} = 4.0~${}mrad, $\sigma_y = 9.0~${}mm, and $\sigma_{\theta y} = 1.1~${}mrad. Additionally, the dispersive angle and position were given a correlation of $\theta_x/x = 0.0741~${}mrad/mm.  The angle and position of the incoming beam were determined from position measurements in the two CRDC detectors upstream of the reaction target.  The beam energy was determined by comparing measured and simulated distributions in the two downstream CRDC detectors for runs where the reaction target was removed. The $^{9}\textrm{Be}(^{29}\textrm{Ne}, ^{27, 28}\textrm{F})$ reactions were treated in the Goldhaber Model \cite{Goldhaber1974306} including a small friction term \cite{Tarasov2004536} to degrade the beam energy by $0.6 \%.$  The transport of charged fragments through the Sweeper was simulated using a third order COSY transformation matrix, produced from measurements of the Sweeper's magnetic field \cite{FrankPhd}.

The resolution of charged particle position and angle measurements was modeled as Gaussian, with $\sigma_\textit{pos} = 1.3$ mm and $\sigma_\textit{ang} = 0.8~${}mrad. These resolutions were determined from data taken with a tungsten mask shadowing the CRDC detectors.  The primary acceptance cut concerning the charged particles was the requirement that they pass through the $\delta = \pm 150$ mm active area of the downstream CRDC.  Neutron time of flight resolution was modeled as Gaussian with $\sigma = 0.3$ ns, and the neutron $x$-position resolution was modeled as a sum of two Laplacian functions:
\begin{equation}
\label{eq:mona_xres}
	p_1 \cdot \frac{e^{- \lvert x/ \sigma_1 \rvert}}{2 \sigma_1} +
	\left(1 - p_1\right) \cdot \frac{e^{- \lvert x/ \sigma_2 \rvert}}{2 \sigma_2} ,
\end{equation}

\noindent with $\sigma_1 = 16.2~\textrm{cm},$ $\sigma_2 = 2.33~\textrm{cm},$ and $p_1 = 53.4 \%.$  The form of \eqnref{eq:mona_xres} and the parameters $\sigma_1,$ $\sigma_2,$ and $p_1$ were determined from shadow bar measurements and GEANT3 simulations \cite{PetersPhd}. As mentioned, the neutron $y$ and $z$ positions were assumed to be at the center of the detection bar, resulting in a uniform uncertainty of $\pm 5$ cm.  The overall resolution and acceptance for the decay of \nuc{28}{F} into $^{27}\textrm{F} + n$ has already been presented in Ref. \cite{PhysRevLett.108.032501}, and the corresponding shapes are essentially identical in the case of $^{27}\textrm{F}^{*}$ breakup.

Due to the low statistics of the present data set, an unbinned maximum likelihood technique was used for parameter estimation \cite{Schmidt1993547}. This technique involves forming a small range, $R_i,$ around each experimental data point and then summing the number of weighted Monte Carlo points that lie within the volume.  To marginalize systematic errors resulting from Monte Carlo fluctuations and non-linearities within the $R_i$, the generated model sets were made large ($\sim 3 \times 10^{6}$ events), and the volume size was chosen to be small ($0.05$ MeV).  

\section{Results and Discussion}

\subsection{$^{27}\textrm{F}$ Excited State}

\begin{figure}
    \centering
    \includegraphics{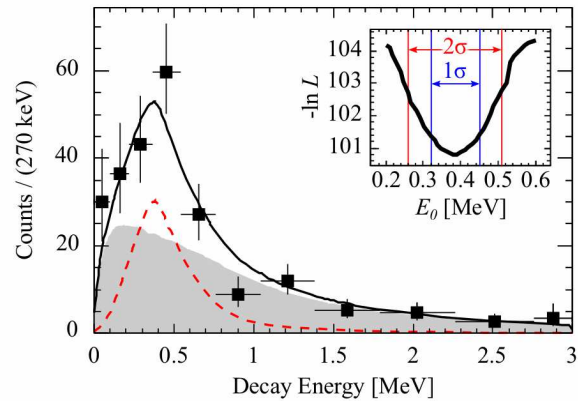}
 	\caption{(color online) Measured relative energy spectrum for $^{26}\textrm{F} + n$ coincidences.  The filled squares with error bars are the experimental data, the dashed red curve is the result of a $380$ keV resonant simulation, the shaded grey curve is a simulation of the Maxwellian non-resonant background ($\Theta = 1.48$ MeV), and the solid black curve is the sum of the resonant and non-resonant models, with a resonant/total fraction of $33\%$. The inset is a plot of the negative log-likelihood as a function of central decay energy, with each point minimized with respect to all other free parameters.}
	\label{fig:f27edecay}
\end{figure}

\begin{figure}
    \centering
    \includegraphics{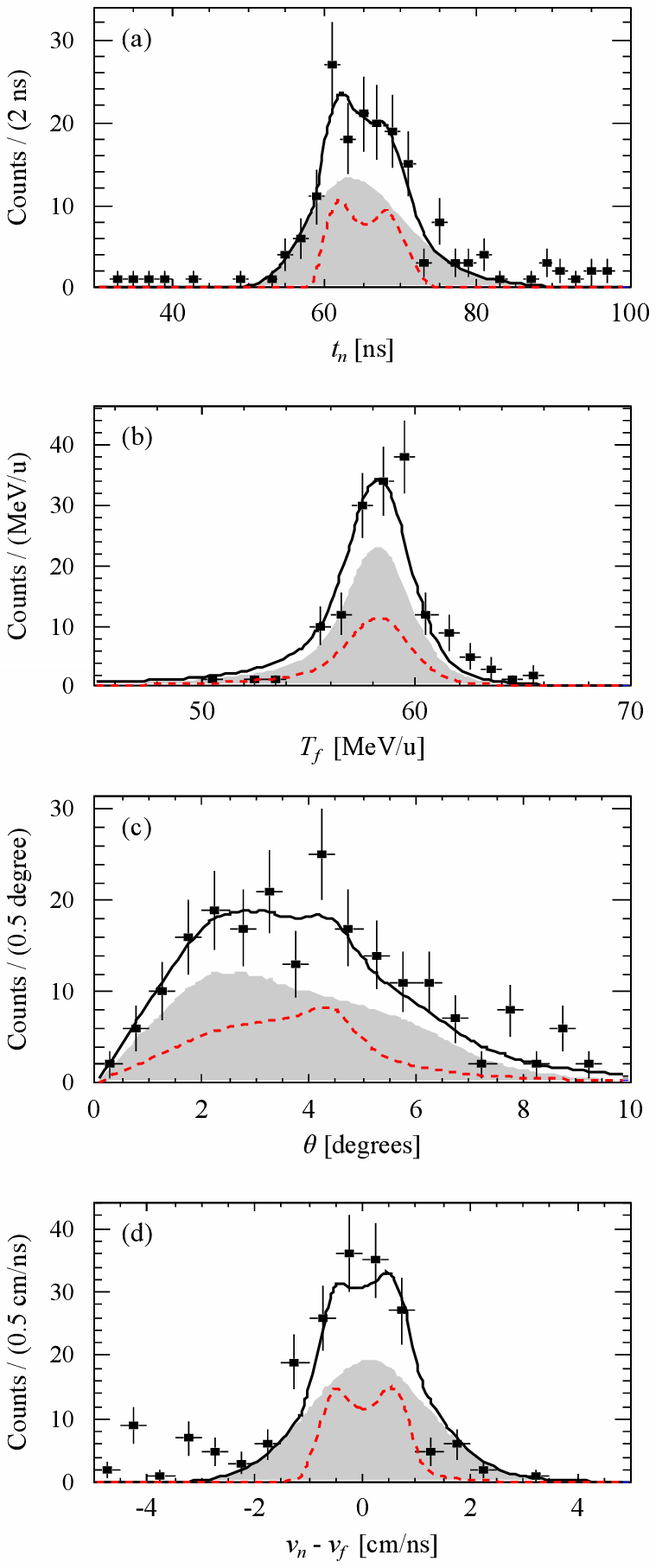}
 	\caption{(color online) Comparison of $^{26}\textrm{F} + n$  simulation and data for a) neutron time of flight; b) charged fragment kinetic energy; c) neutron-fragment opening angle; d) neutron-fragment relative velocity.  The filled black squares are the experimental data, and the solid black, dashed red, and shaded grey curves depict the same simulation components as in \figref{fig:f27edecay}.}
	\label{fig:f27simcompare}
\end{figure}

\begin{figure}
    \centering
    \includegraphics{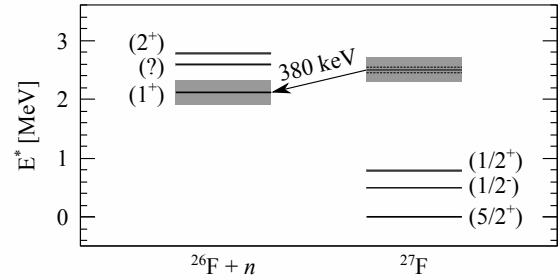}
	\caption{Summary of experimentally known levels in \nuc{26,27}{F}, including the present observation of an unbound excited state in \nuc{27}{F} at $2500$ keV, decaying to the ground state of \nuc{26}{F}. The shaded grey boxes around the various levels indicate the total uncertainty in their placement relative to the \nuc{27}{F} ground state. The dashed lines surrounding the presently observed $2500$ keV level represent the uncertainty on the decay energy only, and the total uncertainty also includes that of the \nuc{27}{F} $1n$ separation energy. All bound excited information is from \cite{Elekes200417}, ground state energies are from \cite{Jurado200743}, and ground state $J^\pi$ are from \cite{Audi19971}. }
	\label{fig:f27levels}
\end{figure}

The black squares in \figref{fig:f27edecay} show the measured decay energy spectrum of $^{26}\textrm{F} + n$ coincidences. As mentioned previously, we expect a non-resonant contribution in the $^{26}\textrm{F} + n$ data, so they were fit with the sum of a Maxwellian distribution and an $\ell = 2$ Breit-Wigner resonance, using the technique outlined in \secref{sec:modelsim}.  In the fit, the resonance energy $E_0,$ resonance width $\Gamma_0,$ Maxwellian temperature $\Theta,$ and resonant/total fraction $f$, were all allowed to vary freely.  In order to extract $E_0,$ the parameter of interest, a profile log-likelihood was constructed by scanning a range of $E_0$ values and plotting the negative log-likelihood $(-\ln[L])$ minimized with respect to the other free parameters ($\Gamma_0,$ $\Theta,$ and $f$).  This profile likelihood curve is displayed in the inset of \figref{fig:f27edecay}, and it reaches a clear minimum at $E_0 = 3\underline{8}0$ keV.  The $n \sigma$ confidence intervals were determined from the $\ln[L_\textit{max}/L] \geq n^2 / 2$ limits. As indicated on the figure, the $1 \sigma$ and $2 \sigma$ confidence intervals were $\pm \underline{6}0$ keV and $^{+1\underline{3}0}_{-1\underline{2}0}$ keV, respectively. The best-fit values of the other parameters were determined to be $\Gamma_0 = \underline{1}0$ keV, $\Theta = 1.48$ MeV, and $f = 33 \%.$  The simulated best fit curves are superimposed on the data in \figref{fig:f27edecay}, with the dashed red curve representing the $380$ keV resonance, the shaded grey curve the Maxwellian background, and the solid black curve their sum.  A comparison between simulation and data is also shown for selected intermediate parameters (neutron time of flight, fragment kinetic energy, neutron-fragment opening angle, and neutron-fragment relative velocity) in \figref{fig:f27simcompare}.

The presumption of $\ell = 2$ decay is based on a pure single-particle model in which the least-bound neutron resides in the $0d_{3/2}$ shell. In reality, configuration mixing and shell evolution could lead to significant contributions from decay with other orbital angular momenta.  Separate analyses using $\ell = 1$ and $\ell = 3$ resonances yield results that do not differ significantly from the $\ell = 2$ case. The lack of sensitivity to $\ell$ values is largely due to experimental resolution, which is limited primarily by uncertainty of the reaction position within the \nuc{9}{Be} target. The width of the measured resonance is almost completely determined by experimental response, overshadowing any differences that might arise from varying the $\ell$ value. Contribution from $\ell = 0$ decays might also be possible, but such decays cannot be separated from the Maxwellian background since the resolved lineshape of the two models is very similar for small absolute scattering lengths ($|a_s| \lesssim 5$ fm). A scattering state near threshold (larger $|a_s|$) is clearly not present since the data display no enhancement at low decay energy.

Only two counts were observed in CAESAR in coincidence with $^{26}\textrm{F} + n$ ($E_{\gamma} = 760$ and $1180$ keV).  In the case of $100\%$ branching to a bound excited state in \nuc{26}{F}, roughly $50$ counts would be expected in CAESAR, based on the approximate $\gamma$-ray detection efficiency of $30 \%$ \cite{Weisshaar2010615}.  Thus the observation of only two $\gamma$ rays in CAESAR indicates that the presently observed decays feed the ground state of \nuc{26}{F}, allowing for an unambiguous assignment of the observed resonance to an excited state in \nuc{27}{F.}  The most recent mass measurements of \nuc{26,27}{F} \cite{Jurado200743} place the \nuc{26}{F} ground state $21\underline{2}0(2\underline{1}0)$ keV above the ground state of \nuc{27}{F}, so we assign the presently observed $3\underline{8}0 (\underline{6}0)$ keV resonance to a $25\underline{0}0 (2\underline{2}0)$ keV excited level in \nuc{27}{F}.  \figref{fig:f27levels} presents this newly observed level along with the other measured states in \nuc{26,27}{F} \cite{Jurado200743, Elekes200417, PhysRevC.85.017303, Audi19971}.

\begin{figure}
    \centering
	\includegraphics{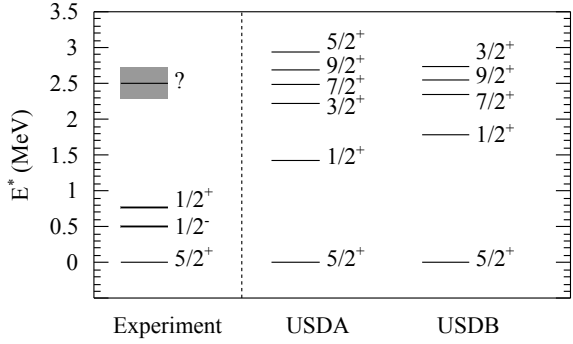}
	\caption{Measured excited levels in \nuc{27}{F} compared to predictions of the USDA and USDB shell models. The first two measured excited states are both from \cite{Elekes200417}, while the state at $2.5$ MeV is from the present work.}
	\label{fig:f27theory}
\end{figure}

To interpret our observations, we have performed shell model calculations using the USDA and USDB interactions \cite{PhysRevC.74.034315}, which operate in the traditional $sd$ model space ($0d_{5/2}$, $1s_{1/2}$, and $0d_{3/2}$ for both protons and neutrons). The calculation results are compared with experiment in \figref{fig:f27theory}.  As seen in the figure, each calculation predicts three or more states in the same energy region as our observation. Extending the calculations to include $pf$ shell components would only complicate the situation since opening up the model space increases the available number of excited state configurations. The assignment of the observed resonance to a specific state is not possible because the reaction ($1p$-$1n$ removal) does not preferentially populate one state over the others.

\subsection{$^{28}\textrm{F}$ Binding Energy}

\begin{figure}
    \centering
    \includegraphics{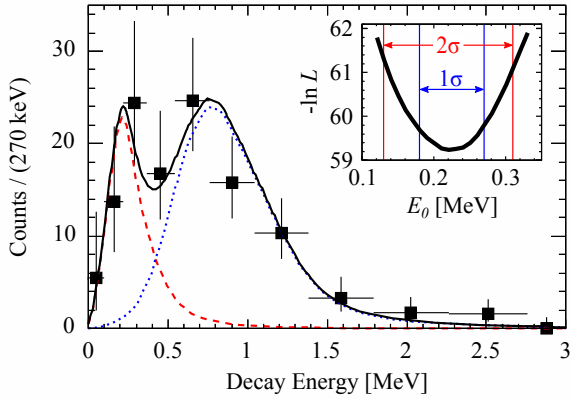}
	\caption{(color online) Relative energy spectrum for $^{27}\textrm{F} + n$ coincidences.  The filled squares with error bars are the experimental data, the dashed red curve is the the $220$ keV resonance simulation, and the dotted blue curve is the $810$ keV simulation.  The solid black curve is the sum of the $220$ keV and $810$ keV resonances, with the relative contribution of the $220$ keV resonance at $28 \%.$ The inset shows the profile log-likelihood as function of the lower resonance energy.}
	\label{fig:f28edecay}
\end{figure}

\begin{figure}
    \centering
    \includegraphics{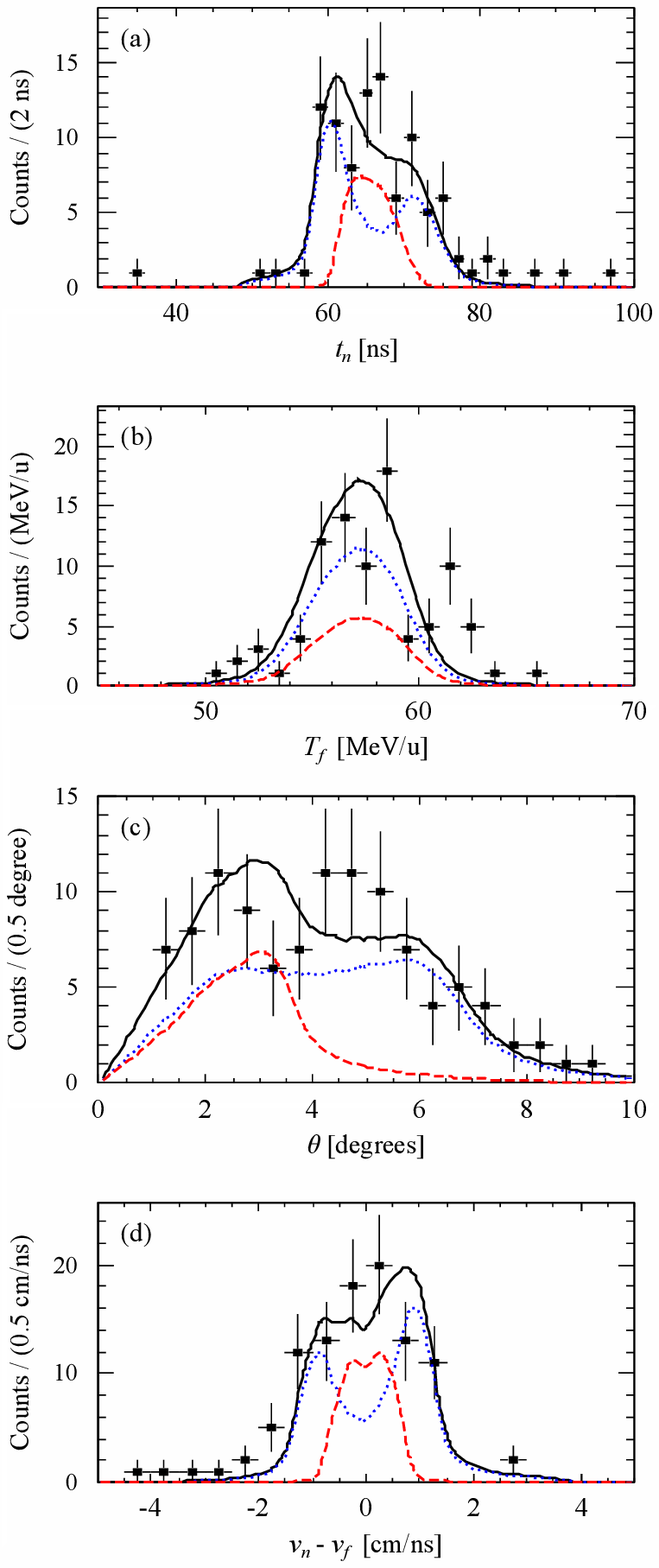}
 	\caption{(color online) Comparison of $^{27}\textrm{F} + n$ simulation and data for a) neutron time of flight; b) charged fragment kinetic energy; c) neutron-fragment opening angle; d) neutron-fragment relative velocity.  The filled black squares are the experimental data, and the solid black, dashed red, and dotted blue curves depict the same simulation components as in \figref{fig:f28edecay}.}
	\label{fig:f28simcompare}
\end{figure}

As discussed in Ref. \cite{PhysRevLett.108.032501}, the measured \nuc{28}{F} decay energy is best described as a sum of two independent $\ell = 2$ Breit-Wigner resonances, with the lower resonance at $2\underline{2}0 (\underline{5}0)$ keV ($\Gamma_0 \equiv 10$ keV), the upper resonance at $810$ keV ($\Gamma_0 \equiv 100$ keV), and the lower resonance composing $28 \%$ of the total area.  As with \nuc{27}{F}, the width of each measured resonance was dominated by experimental resolution, making sensitivity to the resonance $\ell$ value minimal. An $\ell = 0$ scattering state was excluded based on incompatibility with the measured data, and a non-resonant Maxwellian background was not expected since \nuc{28}{F} was populated directly by one-proton knockout from \nuc{29}{Ne}.  The measured \nuc{28}{F} decay energy spectrum is presented in \figref{fig:f28edecay}, along with the best fit two-resonance simulation and the profile log-likelihood curve. Additionally, \figref{fig:f28simcompare} shows a data-simulation comparison for neutron time of flight, fragment kinetic energy, neutron-fragment opening angle, and neutron-fragment relative velocity. No $\gamma$ rays were recorded in CAESAR in coincidence with $^{27}\textrm{F} + n$, and around $30$ would be expected in the case of $100 \%$ branching to excited \nuc{27}{F}. This indicates that the observed resonances feed the \nuc{27}{F} ground state.

\begin{figure}
    \centering
    \includegraphics{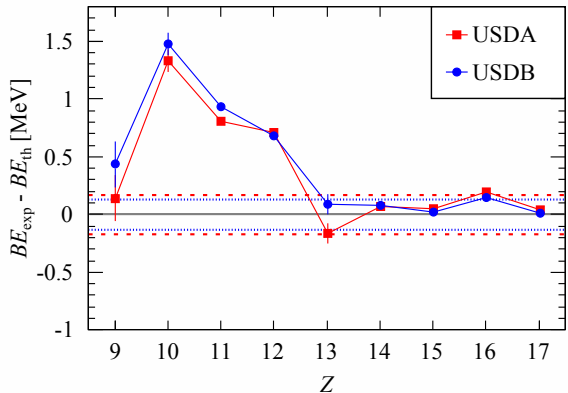}
	\caption{(color online) Difference between experimental and theoretical (USDA, USDB) binding energies for $N=19$ isotones, $9 \leq Z \leq 17.$  The error bars on the data points represent experimental errors only.  The blue dotted and red dashed bands represent the respective $170$ and $130$ keV RMS deviations of USDA and USDB interactions. Experimental values, save for $Z=9$ which is from the present work, are taken from \cite{Jurado200743} if reported there; otherwise they are from the 2003 Atomic Mass Evaluation \cite{Audi2003337}. Figure reproduced from Ref. \cite{PhysRevLett.108.032501}.}
	\label{fig:n19be}
\end{figure}

The present observation of the \nuc{28}{F} ground state as a $2\underline{2}0 (\underline{5}0)$ keV unbound resonance can be combined with the \nuc{27}{F} mass measurement of Ref. \cite{Jurado200743} (\nuc{27}{F} atomic mass excess equal to $246 \underline{3}0 (1 \underline{9}0)~${}keV) to calculate the \nuc{28}{F} binding energy as $1860\underline{4}0 (2\underline{0}0)$ keV.  By comparing measured binding energies with the predictions of the UDSA/USDB shell model, which does not allow for mixing between $sd$  and $pf$ shell configurations, it is possible to qualitatively determine the contribution of $pf$ shell ``intruder'' components in the ground state of a given nucleus. Such a comparison is shown in \figref{fig:n19be} for $N = 19$ isotones with $9 \leq Z \leq 17$.  As seen in the figure, the agreement is very good for the heavier isotones closer to stability $(Z \geq 13)$, while it becomes dramatically worse for the isotones with $10 \leq Z \leq 12$ which lie within the island of inversion.  At $Z = 9$, the good agreement between USDA/USDB and experiment is dramatically recovered, indicating that intruder components play a minimal role in the ground state structure of \nuc{28}{F}. This suggests the existence of a low-$Z$ boundary (or ``shore'') of the island of inversion beginning at $Z = 9.$

\section{Conclusions and Outlook}

In conclusion, we have used the technique of invariant mass spectroscopy to make the first determination of the \nuc{28}{F} binding energy at $1860\underline{4}0 (2\underline{0}0)$ keV. Additionally, we have observed a neutron-unbound excited state in neighboring \nuc{27}{F}, with $25\underline{0}0 (2\underline{2}0)$ keV excitation energy.

Interpretation of the \nuc{27}{F} state in terms of shell model predictions is difficult due to the large number of levels predicted near $2500$ keV and uncertainty in the reaction mechanism used to populate $^{27}\textrm{F}^{*}$.  The level structure of \nuc{27}{F} is relevant to a variety of open questions in nuclear physics, including the transition from pure $sd$ to mixed $sd$-$pf$ neutron configurations and its associated consequences (such as the large oxygen-fluorine dripline shift of six or more neutrons \cite{PhysRevC.64.011301}).  Additionally, it has been suggested \cite{Elekes200417, PhysRevC.83.044304} that proton $p$-$sd$ cross-shell excitations could play a role in the structure of low-lying \nuc{27}{F} excited states, possibly in tandem with $sd$-$fp$ shell breaking on the neutron side. As such, it would be interesting to revisit unbound excited states in \nuc{27}{F} experimentally, using a direct reaction mechanism that can selectively populate specific states. Possible reactions include one- or two-proton knockout (from \nuc{28}{Ne} or \nuc{29}{Na}) and $^{26}\textrm{F} (d,~p)$ in inverse kinematics.

The measured \nuc{28}{F} binding energy indicates a low-$Z$ boundary of the island of inversion at $N = 19$. It would be interesting to further explore this mass region to see if this trend continues.  Extension of the present technique to lighter $N = 19$ isotones ($Z \leq 8$) would be very difficult, if not impossible, since they are all unbound by three or more neutrons \cite{Tarasov199764, Sakurai1999180, Langevin198571, PhysRevC.41.937, PhysRevC.53.647}.  However, a similar technique could potentially be used in the $N = 20$ isotonic chain by performing a direct mass measurement of bound \nuc{29}{F}. For this purpose, the precision obtainable with time-of-flight techniques at current in-flight radioactive beam facilities would likely be sufficient.  Such a measurement would be particularly interesting since the SDPF-M Monte Carlo Shell Model predicts \nuc{29}{F} to have a very large intruder occupation of $91.5 \%$ ($62.7 \%$ two-particle, two-hole excitation and $28.8 \%$ four-particle, four-hole) \cite{PhysRevC.81.041302}.  Measuring its mass would provide the first experimental data on \nuc{29}{F} for comparison with theory and help to better explain the evolution of shell structure in the low-$Z$ ($ < 10$) region around $N = 20$.

\begin{acknowledgments}
The authors thank the NSCL operations
staff for providing a high-quality beam throughout
the experiment and the NSCL design staff for their efforts in the
construction of a magnetic shield for CAESAR. We are also grateful to B. A. Brown
and A. Signoracci for their assistance with shell model calculations.
Finally, we thank the NSCL Gamma Group and MoNA Collaboration for their effort in setting up and supporting the experiment. Funding for this
work was provided by the National Science Foundation under
grants No. PHY-05-55488, No. PHY-05-55439, No. PHY-
06-51627, No. PHY-06-06007, No. PHY-08-55456, and No.
PHY-09-69173.

\end{acknowledgments}

\end{document}